\documentstyle[12pt,aaspp4,psfig,natbib]{article}

\newcommand{\vy}[2]{#1_{\scriptscriptstyle #2}}
\newcommand{\Ly}{Ly$\alpha$}

\def\gtorder{\mathrel{\raise.3ex\hbox{$>$}\mkern-14mu
             \lower0.6ex\hbox{$\sim$}}}
\def\ltorder{\mathrel{\raise.3ex\hbox{$<$}\mkern-14mu
             \lower0.6ex\hbox{$\sim$}}}
\def\proptwid{\mathrel{\raise.3ex\hbox{$\propto$}\mkern-14mu
             \lower0.6ex\hbox{$\sim$}}}
\def\0946{PG~0946+301}

\def\arcsec{\ifmmode '' \else $''$\fi}

\def\arcsecpoint{\ifmmode ''\!. \else $''\!.$\fi}

\def\kms{\ifmmode {\rm km\ s}^{-1} \else km s$^{-1}$\fi}
\def\Msun{\ifmmode {\rm M}_{\odot} \else M$_{\odot}$\fi}
\def\Lsun{\ifmmode {\rm L}_{\odot} \else L$_{\odot}$\fi}
\def\Zsun{\ifmmode {\rm Z}_{\odot} \else Z$_{\odot}$\fi}

\def\ergscm2{ergs\,s$^{-1}$\,cm$^{-2}$}
\def\icm3{{\rm cm}^{-3}}
\def\icm2{{\rm cm}^{-2}}
\def\qo{\ifmmode q_{\rm o} \else $q_{\rm o}$\fi}
\def\Ho{\ifmmode H_{\rm o} \else $H_{\rm o}$\fi}
\def\ho{\ifmmode h_{\rm o} \else $h_{\rm o}$\fi}

\def\vFWHM{\ifmmode v_{\mbox{\tiny FWHM}} \else
            $v_{\mbox{\tiny FWHM}}$\fi}
\def\CCF{\ifmmode F_{\it CCF} \else $F_{\it CCF}$\fi}
\def\ACF{\ifmmode F_{\it ACF} \else $F_{\it ACF}$\fi}
\def\Halpha{\ifmmode {\rm H}\alpha \else H$\alpha$\fi}
\def\Hbeta{\ifmmode {\rm H}\beta \else H$\beta$\fi}
\def\Hgamma{\ifmmode {\rm H}\gamma \else H$\gamma$\fi}
\def\Hdelta{\ifmmode {\rm H}\delta \else H$\delta$\fi}
\def\Lya{\ifmmode {\rm Ly}\alpha \else Ly$\alpha$\fi}
\def\Lyb{\ifmmode {\rm Ly}\beta \else Ly$\beta$\fi}
\def\Lyg{\ifmmode {\rm Ly}\beta \else Ly$\gamma$\fi}
\def\hi{H\,{\sc i}}

\def\ciii{\ifmmode {\rm C}\,{\sc iii} \else C\,{\sc iii}\fi}
\def\civ{\ifmmode {\rm C}\,{\sc iv} \else C\,{\sc iv}\fi}

\def\nv{N\,{\sc v}}

\def\o5007{[O\,{\sc iii}]\,$\lambda5007$}

\def\ovi{O\,{\sc vi}}
\def\ovii{O\,{\sc vii}}

\def\o{\o}
\begin{document}

\title{CONTRASTING THE UV AND X-RAY \ovi\ COLUMN DENSITY INFERRED FOR
THE OUTFLOW IN NGC~5548 }


\author{
Nahum Arav\altaffilmark{1}, 
Jelle  Kaastra\altaffilmark{2},
Katrien Steenbrugge\altaffilmark{2}, 
Bert Brinkman\altaffilmark{2},
Rick Edelson\altaffilmark{3},
Kirk~T.~Korista\altaffilmark{4},
Martijn de~Kool\altaffilmark{5} }

\altaffiltext{1}{CASA, University of Colorado, 389 UCB, Boulder, CO 80309-0389,
I:arav@colorado.edu}
\altaffiltext{2}{SRON National Institute for Space Research
Sorbonnelaan 2, 3584 CA Utrecht, The Nether\-lands}
\altaffiltext{3}{Astronomy Department, UCLA}
\altaffiltext{4}{Western Michigan Univ., Dept.\ of Physics, 
Kalamazoo, MI 49008-5252}
\altaffiltext{5}{Research School of Astronomy and Astrophysics, ANU ACT,
 Australia}

\begin{abstract}

We compare X-ray and UV spectroscopic observations of NGC~5548. Both
data sets show \ovi\ absorption troughs associated with the AGN
outflow from this galaxy.  We find that the robust lower limit
on the column density of the \ovi\ X-ray trough is seven times larger
than the column density found in a study of the \ovi\ UV troughs. This
discrepancy suggests that column densities inferred for UV troughs of
Seyfert outflows are often severely underestimated. We identify the
physical limitations of the UV Gaussian modeling as the probable
explanation of the \ovi\ column density discrepancy. Specifically,
Gaussian modeling cannot account for a velocity dependent covering
fraction, and it is a poor representation for absorption associated
with a dynamical outflow.  Analysis techniques that use a single
covering fraction value for each absorption component suffer from
similar limitations. We conclude by suggesting ways to improve the UV
analysis.

\end{abstract}

\section{INTRODUCTION}

Outflows in Seyfert galaxies are evident by resonance line absorption
troughs, which are blueshifted with respect to the systemic redshift
of their emission counterparts. Velocities of several hundred \kms\
(Crenshaw et~al.\ 1999; Kriss et~al.\ 2000) are observed in both
UV resonance lines (e.g., \civ~$\lambda\lambda$1548.20,1550.77,
\nv~$\lambda\lambda$1238.82,1242.80, \ovi~$\lambda\lambda$1031.93,1037.62
and \Ly), as well as in X-ray resonance lines (Kaastra et~al.\ 2000;
Kaspi et~al.\ 2000). Similar outflows (often with significantly higher
velocities) are seen in quasars which are the luminous relatives
of Seyfert galaxies (Weymann et~al.\ 1991; Korista, Voit, Morris, \&
Weymann 1993; Arav et~al.\ 2001a). Reliable measurement of the absorption
column densities in the troughs are crucial for determining the ionization
equilibrium and abundances of the outflows, and the relationship between
the UV and the ionized X-ray absorbers.

NGC~5548 is one of the most studied Seyfert galaxies, including intensive
reverberation campaigns (Netzer \& Maoz 1990; Clavel et al.\ 1991;
Peterson et~al.\ 1991; Korista et~al.\ 1995), line studies (Krolik
et al.\ 1991; Ferland et al.\ 1992; Goad \& Koratkar 1998; Kaspi \&
Netzer 1999; Korista \& Goad 2000), and theoretical modeling (Done \&
Krolik 1996; Bottorff, Korista \& Shlosman 2000; Srianand 2000). The
intrinsic absorber in NGC~5548 was studied in the UV using the {\em
International Ultraviolet Explorer} (Shull \& Sachs 1993), the {\em HST}
Goddard High Resolution Spectrograph (GHRS) (Mathur, Elvis \& Wilkes
1999) and Space Telescope Imaging Spectrograph (STIS) (Crenshaw \&
Kraemer 1999), the {\em Far Ultraviolet Spectroscopic Explorer} ({\em
FUSE}) (Brotherton et~al.\ 2002), and in X-ray with the {\em ASCA}
(George et~al.\ 1998) and {\em Chandra} (Kaastra et~al.\ 2000, 2002)
satellites. These high quality observations combined with the relative
simplicity of its intrinsic absorption features make NGC~5548 an excellent
target for understanding the nature of Seyfert outflows.

In the last few years our group (Arav 1997; Arav et~al.\ 1999a; Arav
et~al.\ 1999b; de~Kool et~al.\ 2001; Arav et~al.\ 2001) and others
(Barlow 1997, Telfer et~al.\ 1998, Churchill et~al.\ 1999, Ganguly
et~al.\ 1999) have shown that in quasar outflows most lines are
saturated even when not black. We have also shown that in many cases
the shapes of the troughs are almost entirely due to changes in the
line of sight covering as a function of velocity, rather than to
differences in optical depth (Arav et~al.\ 1999b; de~Kool et~al.\
2001; Arav et~al.\ 2001a). Gabel et al.\ (2002) show the same effect in
the outflow troughs of NGC~3783. As a consequence, the column
densities inferred from the depths of the troughs are only lower
limits.

These results have led us to suspect that the current determination of
column densities in Seyfert outflows is highly uncertain. Recently,
Arav, Korista \& de~Kool (2002) have re-analyzed the {\em HST} high
resolution spectroscopic data of the intrinsic absorber in NGC~5548
and found that the \civ\ absorption column density in the main trough
is at least four times larger than previously determined. Furthermore,
in the same paper it was shown that similar to the case in quasars,
the shape of the main trough is almost entirely due to changes in
covering fraction as a function of velocity, and is not due to
varying column density as a function of velocity.

\pagebreak

An important new saturation diagnostic is available in the
form of UV and X-ray absorption troughs that arise from the same
ion. The relationship between the ionic column density ($\rm{N_{ion}}$) 
and the optical depth ($\tau$) of an absorption trough is given by
(see eq.\ (3) in Arav et al.\ 1999a):
\begin{equation} \rm{N_{ion}}  = \frac{ 3.8 \times 10^{14}~\rm{cm}^{-2} }{
\lambda_{\rm{o}}f_{ik} } \int {\tau\/(v) dv}, 
\end{equation}
where $\lambda_{\rm{o}}$ and $f_{ik}$ are the wavelength and
oscillator strength of the transition, respectively, and $v$ is the
velocity in \kms.  Using the apparent optical depth ($I=e^{-\tau}$,
where $I$ is the residual intensity) in equation (1), a trough with
the same $I(v)$ yields fifty times higher $\rm{N_{ion}}$ for an X-ray
line as compared to a UV line with the same $f_{ik}$, simply because
its wavelength is fifty times smaller.  In cases where the UV trough
is saturated and its shape is determined by changes in covering
fraction as a function of velocity, it is virtually impossible to
distinguish between $\tau$ values above $\sim3$ (see Arav, Korista \&
de~Kool 2002).  For equal oscillator strength, the X-ray trough of the
same ion will have $\tau=1$ when the UV trough reaches
$\vy{\tau}{UV}=50$. Therefore, the X-ray trough is sensitive to a much
higher column density than the UV one and can provide a crucial
saturation diagnostic for the UV troughs.

In this paper we compare the \ovi\ column density inferred
independently from UV and X-ray absorption features in the spectrum of
NGC~5548, and demonstrate the large discrepancy between the two
values. The X-ray derived \ovi\ column density is seven times larger
than the UV one.  Since the X-ray determination is a robust lower
limit, we conclude that the UV lines are much more saturated than
originally inferred, and identify the UV analysis as the probable
cause for this discrepancy.  Specifically, Gaussian modeling cannot
account for a velocity dependent covering fraction, and more generally
Gaussian modeling is a poor representation for absorption associated
with a dynamical outflow.

\section{CONTRASTING THE UV AND X-RAY INFERRED \ovi\ COLUMN DENSITY}

\subsection{Data Description}

NGC 5548 was observed by {\em FUSE} on 2000 June 7 with a total exposure
time of 25 ksec. A full description of the observations, reduction process
and plots of the spectra are presented in Brotherton et~al.\ (2002). These
observations include the spectral region associated with both the emission
and absorption features of the \ovi~$\lambda\lambda$1032,1038 doublet
in the rest frame of NGC~5548.

Our {\em XMM-Newton} data were taken on 2001 July 12. The total
exposure was 116 ksec and the data were reduced using the standard SAS
analysis package. Here we only discuss the Reflection Grating
Spectrometer (RGS) spectra. A detailed description of the RGS
instruments is given by Den Herder et~al.\ (2001). Briefly, there are
two almost identical RGS detectors aboard {\em XMM-Newton} that cover
the 5$-$38~\AA\ range with an almost uniform spectral resolution of
0.05~\AA, and have a maximum effective area of about 140~cm$^2$ at
15~\AA. In the present work we focus upon the \ovi\ line region around
22~\AA. Due to failure of one of the 9 CCD's of RGS2, only data from
RGS1 are available in this wavelength band and analyzed here.  The
spectrum was fitted using a continuum plus absorption components as
described by Kaastra et~al.\ (2002).  A full analysis of the spectrum
will be given elsewhere (Steenbrugge et~al., 2002). as well as timing
analysis of the data (Markowitz et al. 2002) and
analysis of the European Photon Imaging Camera (EPIC) data
(Pounds et al. 2003).

\subsection{\ovi\ Column Density Extractions}

In the {\em XMM-Newton}/RGS spectrum there is a distinct absorption
line visible at a rest wavelength of 22.0~\AA\ due to \ovi\ (see
Fig.\ 1).  The line occurs in the wavelength range of the well-known
He-like triplet of \ovii, from which the forbidden line at 22.1~\AA\
and the intercombination line at 21.8~\AA\ are seen in emission, and
the resonance line at 21.6~\AA\ in absorption (cf.\ Kaastra et~al.\
2002). The \ovi\ absorption line at 22.0~\AA\ is in fact a satellite
to these He-like lines, and its importance for AGN spectra was first
recognized by Pradhan (2001). The line (in fact an unresolved blend of the
so-called $r$ and $u$ lines) is caused by inner-shell photo-excitation
of an \ovi\ ion in the ground state to the 1s2p($^1$P)2s\,$^2$P$_{1/2}$
and 1s2p($^1$P)2s\,$^4$P$_{3/2}$ levels, respectively. We used atomic
data for this transition as calculated by Behar (2002, in preparation).
The wavelength for the blend of the two components is 22.007~\AA\, and
the combined oscillator strength $f$ is 0.525 (Pradhan gives 22.05~\AA\
and $f=0.576$).

The equivalent width of this \ovi\ trough is 40 $\pm10$~mA, and
therefore a {\it lower limit} for the associated column density is
$3.2\pm0.8\times10^{16}$ cm$^{-2}$, where the error is directly associated
with the signal-to-noise of the data. A detailed description of the fitting 
method used to extract this \ovi\ column density value is given
in Steenbrugge et~al.\ (2002)

We stress that this is a lower
limit since: a) There may be unresolved absorption components due to the
$R=500$ resolution of the RGS at this wavelength. b) The trough may
show non-black saturation even for resolved components, similar to the
situation evident in the UV troughs associated with the outflow.
c)  The adjacent \ovii\  f emission line, can partly fill in the \ovi\ trough.
We note that the FWHM of the \ovi\ trough is 800$\pm$200 \kms,
and that it's centroid velocity is $-370\pm160$ \kms\ for a systemic redshift 
of $z=0.017175$ (see redshift discussion in Brotherton et~al, 2002).

\begin{figure}
\centerline{\psfig{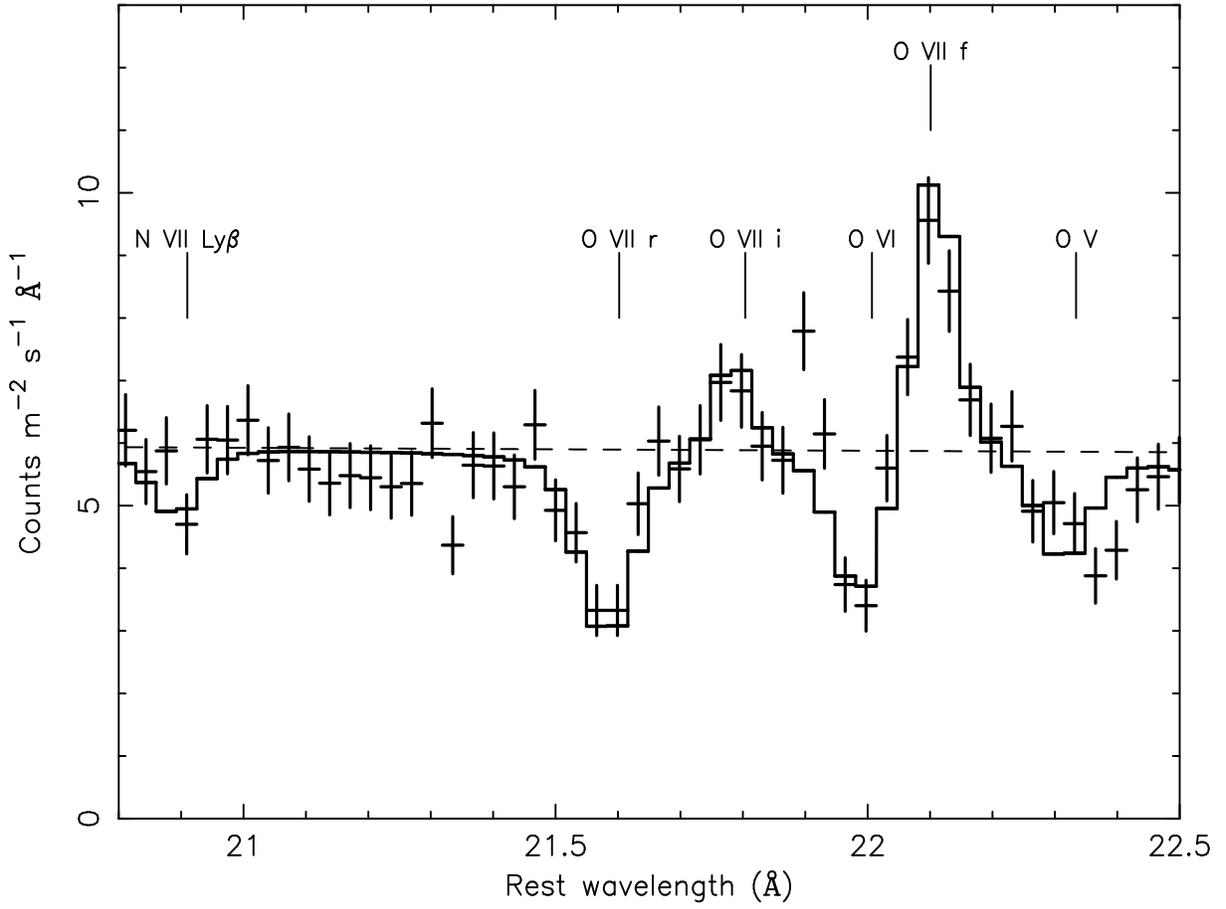}}
\caption{{\em XMM}/RGS data of NGC~5548. The absorption trough at 22.0\AA\
is due to the \ovi\ transitions discussed in the text.} 
\label{fig_xmm}
\end{figure}

Brotherton et~al.\ (2002) analyzed the {\em FUSE} spectra of NGC~5548
in order to determine the \ovi\ absorption column density from the 
\ovi~$\lambda\lambda$1031.93,1037.62 (oscillator strength of 0.13 and
0.066, respectively; Verner, Verner \& Ferland 1996). They used
the program SPECFIT (Kriss 1994) to fit the \ovi\ spectral region using
the following assumptions:  Approximating the intrinsic absorption as
several components with Gaussian distributions in optical depth that are
permitted to partially cover the background source, where the optical depth
ratio of the doublets is constrained to be 2:1 at all velocities. They
have looked at two types of models. one where the outflow covers all
the emission components (continuum, broad emission lines, and narrow
emission lines [NELs]), and one where the flow does not cover the NELs.
Since the model with un-covered NELs gives higher $N_{\ovi}$ estimates,
we will concentrate on it, noting that our discussion is similarly
applicable to the other model.

The overall quality of the Brotherton et~al.\ (2002) fit is very good
(see their Fig.~3) and the statistical error quoted for the $N_{\ovi}$
associated with each of the absorption components is mostly between
10--20\% (their Table 2).  We note though, that their model includes
43 free parameters: Eight absorption Gaussian with four free
parameters each (position, depth, width and covering fraction), three
emission Gaussians with three free parameters each (position, hight
and width), and two free parameters for the continuum (flux and
slope). Although the goodness of the fit is impressive, it is not
surprising given the number of free parameters involved.  Summing up
the $N_{\ovi}$ determinations in each component we obtain a total
$N_{\ovi}=4.9\pm0.6\times10^{15}$ cm$^{-2}$, where for a conservative
error estimate we simply added the quoted individual errors. The total
width of the \ovi\ UV absorption complex is $\sim1200$ \kms, which is
consistent with the width of the \ovi\ X-ray trough (see above).

We therefore possess two very different estimates for the \ovi\
absorption column density associated with the outflow in NGC~5548.
On the one hand the {\em XMM}/RGS observation gives a lower limit
$N_{\ovi}\geq3.2\pm0.8\times10^{16}$ cm$^{-2}$, on the other hand a Gaussian
fitting of the {\em FUSE} data yields $N_{\ovi}=4.9\pm0.6\times10^{15}$
cm$^{-2}$. 

\subsection{Kinematic Distribution of the X-ray-Infered \ovi\ Column Density}

At 22 \AA\ the resolution element of the RGS instrument is roughly 700
\kms.  In order to obtain the deep \ovi\ absorption trough seen in the data
(Fig.\ 1), the absorbing material must therefore be spread over a
similar velocity interval.  An absorption component with a much
narrower velocity width cannot account for the X-ray trough even if
its optical depth is very high.  This narrow yet deep intrinsic trough
will be transformed into a shallow trough of roughly 700 \kms\ width
by the line spread function of the instrument.

Brotherton et~al.\ (2002) measured FWHM$<150$ \kms\ for UV components
2-6 of the outflow.  Therefore, the high X-ray-inferred $N_{\ovi}$
cannot be hidden in only one, two, or three of these components even if
they are highly saturated. UV components 0.5 and 1 are wider (350 and
280 \kms, respectively). However, even combined, these components
cannot by themselves account for the X-ray-inferred $N_{\ovi}$. Their
combined width and shape are not enough to produce the observed X-ray
trough even if we assume both components are highly saturated.  We
conclude that since the X-ray-inferred $N_{\ovi}$ must be associated
with the UV \ovi\ kinematic components, at least half of these
components have to be highly saturated to allow for the observed X-ray
\ovi\ absorption trough.

\section{WHAT CAUSES THE DISCREPANCY IN THE \ovi\ COLUMN DENSITY DETERMINATIONS?}

\subsection{The Plausibility of Real Temporal Changes}

A difference of 13 months exist between the {\em FUSE} (2000 June 7)
and {\em XMM} (2001 July 12) observations of NGC~5548. It is therefore
important to address the possibility that the order of magnitude
difference in inferred $N_{\ovi}$ is caused by temporal changes.
In table 1 we list the relevant UV and X-ray observations, including
the  continuum flux associated with each.
Based on the four available epochs of high-resolution UV spectroscopy
of the object, we conclude that strong variations in the shape and
strength of the UV troughs over a year time-scale are
unlikely. Components 2, 4 and 5 of the \civ\ trough (following
Crenshaw \& Kraemer 1999) show only marginal changes in the three
available epochs of high-resolution spectroscopy: {\em HST}/GHRS
spectrum taken in 1996 August 24 (Mathur, Elvis \& Wilkes 1999), {\em
HST}/STIS observation from 1998 March 11 (Crenshaw \& Kraemer 1999)
and {\em HST}/STIS observation from 2002 January 22 (Crenshaw et~al.\
in preparation). Only components 1 and 3 (which are blended to some
extent) show significant variation (at the $\sim$50\% level).  The
main \civ\ trough (component 4) showed less than 10\% changes in
residual intensity between all three epochs.  Furthermore, component 4
in the \civ\ trough is almost identical in shape and depth to the main
component in the \ovi\ trough covered by the {\em FUSE} observations
(see Fig.~4 in Arav et~al.\ 2002).  Such similarity argues for both
small temporal variation in the troughs, as well as for a moderate to
high degree of saturation in both the \civ\ and \ovi\ troughs (see
discussion in Arav et~al.\ 2002).  These comparisons suggest that the
shape of the main UV troughs in NGC~5548 change rather moderately on a
time scale of 2 -- 6 years. 

We note that if the UV troughs show non-black saturation, large column
density changes are possible even when the absorption troughs do not
change appreciably.  However, UV analysis techniques (Gaussian
modeling or otherwise) have a limited dynamical range for inferring
real optical depth (see \S~3.3). Therefore, if the troughs' shape did
not change appreciably it is very difficult to deduce a real change in
the \ovi\ absorbing column density even if the level of saturation
increased considerably. Furthermore, the X-ray data suggest that
dramatic $N_{\ovi}$ changes are unlikely over a one year time scale.

Recently (January 2002) a very long {\em Chandra} LETGS observation
(340 ksec) of NGC~5548 was obtained (Kaastra et~al.\ in
preparation). Preliminary analysis indicates that the detected \ovi\
trough is very similar to the one seen in the {\em XMM}/RGS
observation.  Such occurrence suggests that an order of magnitude
changes in the EW of the \ovi\ X-ray trough between the {\em FUSE}
(2000 June 7) and {\em XMM} (2001 July 12) observations of NGC~5548 is
unlikely.  There is also an earlier epoch {\em Chandra} LETGS
observation taken on December 1999 (86 ksec; Kaastra et~al.\ 2000) and
have now been reanalyzed using improved wavelength and effective area
calibration by Kaastra et~al.\ (2002). The S/N of this observation is
significantly lower than the S/N of either the {\em XMM} data (shown
here), or the new {\em Chandra} LETGS observation. The \ovi\ line is
marginally detected (Fig.~3 in Kaastra et~al.\ 2000), with a 1$\sigma$
upper limit of $3.4\times10^{16}$~cm$^{-2}$, which is consistent with
the {\em XMM} measurement. 

Additional support for relatively small changes in the \ovi\ X-ray
trough on a year timescale comes from comparing total fluxes in the
2-10 keV bands.  The X-ray flux dropped by roughly 50\% between the
two higher quality observations ({\em XMM} and January 2002 {\em
Chandra}), yet the derived $N_{\ovi}$ values are similar in both
observations.  This occurence suggest that the $N_{\ovi}$ value is
quite insensitive to considerable flux changes over a year time scale.

\begin{table}[htb]
\begin{center}
\begin{tabular}{llll}
\multicolumn{4}{c}{\sc Table 1: NGC~5548 Continuum Fluxes In Different Epochs} \\
\hline
\hline
Epoch   &        Instrument &  Flux  & Reference \\
\hline
August 1996 &  {\em HST}/GHRS   &   $5.6\times10^{-14,a}$ & Crenshaw et~al.\ 2003 \\
March 1998  &  {\em HST}/STIS   &   $6.6\times10^{-14,a}$ & Crenshaw et~al.\ 2003 \\
December 1999 & {\em Chandra}/LETGS & $4.0\times10^{-11,b}$ &   Kaastra et~al.\ 2002 \\
June 2000   &  {\em FUSE}       &   $1.5\times10^{-14,c}$ & Brotherton et~al.\ 2002 \\
July 2001   &  {\em XMM}/RGS    &    $5.0\times10^{-11,b}$   & Pounds et al. 2003 \\
January 2002 & {\em Chandra}/LETGS & $2.6\times10^{-11,b}$ & Kaastra et~al.\ 2003 \\
January 2002 & {\em HST}/STIS   &   $2.0\times10^{-14,a}$ & Crenshaw et~al.\ 2003 \\
\hline
\end{tabular}
\end{center}
$^a$ - $F_{\lambda}$ at 1360 \AA\ in units of ergs s$^{-1}$ cm$^{-2}$ \AA$^{-1}$ \\
 $^b$ - Flux in the 2-10 keV band in units of  ergs s$^{-1}$ cm$^{-2}$ \\ 
 $^c$ - Avarage $F_{\lambda}$ between 1110--1150 \AA\ in units of ergs s$^{-1}$ 
   cm$^{-2}$ \AA$^{-1}$ \\
\end{table}

\subsection{Validity of the Underlying Physical Model}

The modeling of the absorption troughs presented in Brotherton et~al.\
(2002) is the most sophisticated and detailed Gaussian fitting of an
AGN outflow to date.  As mentioned above, the ``goodness'' of their
fit is impressive.  It is clear that a Gaussian fitting algorithm can
obtain a low $\chi^2$ fit between the data and a model
spectrum. However, after obtaining a good fit the important question
is how physically valid is the Gaussian fitting process?  In other
words, how well would the absorption column densities determined from
Gaussian components represent the actual column densities of the AGN
outflow?

Gaussian fitting is a common analysis tool for spectral absorption features.
In particularly it is used in ISM and IGM studies. In these environments
a Gaussian distribution of optical depth as a function of velocity seems
physically plausible. This is because there is little dynamics involved,
and therefore a thermal velocity distribution function yields a Gaussian
optical depth function (even if ``turbulent'' broadening is invoked to
explain the larger-than-thermal width). Furthermore, for ISM and IGM
the geometrical size of the absorption clouds are much larger than the
emission sources they cover. Therefore, the assumption that along our
line of sight the absorber covers the emission source completely and
uniformly is  very reasonable.

When these models are adapted to represent AGN outflows two
ingredients are added: a constant covering fraction to allow for
partial line-of-sight-covering of the emission source, and a Gaussian
velocity width much larger than is found for ISM and IGM clouds (in
order to account for the width of the outflow's absorption
features). The physical applicability of both ingredients is highly
questionable: First, careful analysis of high-resolution, high S/N
spectra of AGN outflows reveals that the covering fraction is a strong
function of velocity (Arav et~al.\ 1999b; de~Kool et~al.\ 2001, Gabel
et al.\ 2002). In particular, this is also the situation we find for
the \civ\ trough of NGC~5548 (Arav et~al.\ 2002). The strong
dependence of the covering fraction upon velocity naturally undermines
the constant covering fraction assumption.

Similarly, the large Gaussian velocity width invoked to fit the
outflows features is also problematic. For acceptable photoionization
equilibria, clouds' thermal widths are $\ltorder20$ \kms. For IGM
clouds a ``turbulent'' broadening is invoked to explain the
larger-than-thermal width. However, the FWHM of this broadening is
rather small (typical values for the Doppler parameter are $b\approx$
16 -- 46 \kms; Sembach et al. 2001) and is associated with motions on
scales of hundreds of kpc. In contrast the FWHM of Gaussian absorption
components in the Brotherton et~al.\ (2002) modeling reaches 350 \kms\
and is associated with length scales of several parsecs at most. The
larger FWHM and the much smaller length scales associated with the AGN
outflow weakens the physical plausibility of Gaussian optical depth
distribution as compared with the IGM case. We conclude that there is
little physical basis for modeling outflow troughs as Gaussians clouds
with a constant covering fraction.

Both the strong dependence of the covering fraction upon velocity and the
large width of the outflow absorption features can be simply accounted
for by appealing to the dynamical nature of the outflow. Whereas it is
reasonable to expect that a cloud will have a single covering factor
(although, see de~Kool et~al.\ 2002), this is not the situation for an
outflow, where the covering factor can be a strong function of velocity.
Toy models which illustrate these kinematic/geometric effects are found
in Arav (1996) and Arav et~al.\ (1999a).  A similar picture arises in
global models for the structure of AGN (Elvis 2000; Elvis 2001).

\subsection{Asymmetric Errors Associated with the Extraction of Column Densities}

Another important issue is the estimate of errors associated with the
extraction of column densities from the absorption troughs. A formal
statistical error from minimizing the $\chi^2$ between the data and a
model spectrum based on a prescribed $\tau(v)$
can greatly underestimate the real column density
errors.  For example, let us assume that
at a given point the covering fraction is exactly 0.5. In this case
$I=0.5+e^{-\tau}$, where $I$ is the residual intensity and $\tau$ is
the optical depth. Assuming $I=0.531$, we obtain from the above
relationship $\tau=3.5$.  However, even a very small $\Delta I=0.03$
error will cause the following error in optical depth,
$\tau=3.5^{+3.9}_{-0.8}$, and of course if $\Delta I=0.031$ the formal
error will be $\tau=3.5^{+\infty}_{-0.8}$. An error of $\Delta I=.03$
corresponds to a S/N=30, which is rarely (if ever) achieved for high
resolution UV spectroscopy of AGN outflows. We therefore conclude that
when the derived optical depth value is above 3 it is virtually
impossible to set a meaningful positive error estimate. In such cases
we should simply acknowledge that the derived optical depth value is
{\it a lower limit}. 

One can argue that a proper error analysis based on minimizing
$\chi^2$ should also yield the $\tau=3.5^{+\infty}_{-0.8}$ result.
However, this is only correct for one resolution element.  Once we use
a prescribed $\tau(v)$ to fit a full trough, we lose our ability to
distinguish between highly saturated regions in the trough and less
saturated ones. The formal $\chi^2$ $\tau$ error will be obtained from
averaging over the entire trough, whereas in some parts of the trough
the actual error can be very large and dominate the derived column
density error. This situation is accentuated when the shape of the
trough is dominated by velocity dependent covering fraction and not by
$\tau(v)$ variations.

For example, absorption component 4 in the
Brotherton et~al.\ (2002) modeling has a maximum optical depth of 7 in
the blue doublet component (extracted from the parameters in their
Table 2 and using their equations (1) and (2)), and therefore a
maximum optical depth of 3.5 in the red doublet component. Based on
the analysis presented above, it is possible that the real optical
depth in this absorption component is far larger.

\pagebreak

\subsection{Different continuum emitting regions in the X-ray and UV}

For \ovi\ the relevant UV continuum is near 1030 \AA, whereas that of
the X-ray one is near 20 \AA. It is plausible that the size and/or
location of the emitting region at these wavelengths is quite
different.  It is possible that the outflow may cover different
fractions of the continuum source, depending upon the wavelength.
Such differences will complicate the comparison between UV and X-ray
determined absorption column density of the same ion. Unfortunately,
little is yet understood about the continuum emitting mechanism(s) in
AGN. Thus, we cannot quantify the effects expected from the difference
in the respective emitting regions.  

However, in the context of the current investigation the important
question is whether size/location differences in the emitting regions
can explain the discrepancy between the UV and X-ray inferred \ovi\
column densities.  Two combined characteristics of
the NGC 5548 \ovi\ absorber leads us to conclude that such an outcome
is not plausible.  First, the velocity width and the overall depth of
the \ovi\ UV and X-ray absorption troughs are quite similar. Second,
we are comparing an apparent \ovi\ X-ray column density (derived from
Eq.\ (1) using the apparent optical depth, i.e., assuming no
saturation) with  the inferred real UV
\ovi\ column density (which should account for saturation effects). It
is easy to hide a large column density in a saturated trough
(saturation in partially covering components or in unresolved
components), but it is difficult to decrease the lower column density
limit for a given trough's shape. For example, if the covering
fraction of the X-ray trough is much larger than the UV one, then the
apparent X-ray column is larger than the apparent UV column.  However,
in that case we should observe a much deeper X-ray trough compared with
the UV one, which is not seen in the NGC 5548 data.  We conclude that
it is unlikely that different UV and X-ray emitting region geometry
will allow the apparent X-ray column density to be an order of
magnitude larger than the real UV column density.  Moreover, the
striking kinematic similarity of the \Lyb\ UV trough and the \ovii\
X-ray troughs in NGC~3783 (see Fig.\ 2 in Gabel et al.\ 2003) strongly
suggests that a large overlap exists between the X-ray and UV
continuum emitting regions along our outflow-intersecting line of
sight.

We note though, that the argument can be turned around.  A detailed
comparison of X-ray and UV absorption troughs rising from the same ion
may give us an important handle on the geometries of these different
emitting regions.  However, such a comparison necessitates higher
quality X-ray data than those described here.

\section{DISCUSSION}

The \ovi\ data in the NGC~5548 {\em FUSE} observation have a
S/N$\approx$10 (Brotherton et~al.\ 2002). Such limited S/N data do not
allow a direct determination of the covering fraction and optical
depth as a function of velocity. This forces the modeler to use some
combination of physically weak assumptions, in descending order of
``weakness'' these are:

\begin{enumerate}
\item{The apparent optical depths are equal to the real ones.}
\item{The optical depth is assigned a prescribed function (normally
a Gaussian).}
\item{Each component of the flow is given a single covering fraction
value, ignoring the possibility of velocity dependent covering.}
\end{enumerate}
Higher S/N observations are therefore needed in order avoid the practical
necessity of using these assumptions.

We point out that a careful solution of the optical depth and covering
fraction as a function of velocity (using doublet lines, e.g., Arav
et al.\ 2002) can greatly increase the dynamical range for $\tau$
determination.  Using the example from \S~3.3, an $I=0.531$ gives
$\tau_{apparent}=0.6$. Solving for $\tau$ using the doublet equations
can yield robust estimates up to $\tau=3$ in the weak doublet component,
or $\tau=6$ in the strong one, which increases our dynamical range for
$\tau$ determination by an order of magnitude. Furthermore, even in cases
of higher saturation, lower limits for $\tau$ and hence for $N_{ion}$
are  useful in constraining the physical conditions of the absorber,
provided we acknowledge that these are indeed lower limits (see the case
of BALQSO PG~0946+301; Arav et~al.\ 2001b).

Finally, in view of the caveats described in this paper, we advocate
the following cautious recipe for determining column densities from
absorption troughs of AGN outflows:

\noindent a) For a singlet line, with no other observed lines from the
same ion, we can only measure the apparent optical depth as a function
of velocity and use the resultant integrated column density as {\it a
lower limit}. This is because it is conceptually impossible to decouple
the effects of optical depth from those of covering factor for a single
line, In particular, this is the case for \Ly\ in the {\em HST}/STIS band.

\noindent b) For a doublet line we can solve for both the optical depth
and the covering fraction as a function of velocity. For this we need
a sufficiently narrow absorption complex so that absorption features
from the two doublet component do not blend. If blending occurs we
have to resort to apparent optical depth estimates. Furthermore, for
the unblended case, if the derived optical depth values are larger than
$\sim$3 for the weak doublet component, we have to establish a realistic
lower limit (based on the S/N and emission models uncertainties), which
in practice will never be much higher than $\tau\sim$3.

\noindent c) A line series from the same ion offer the best column density
diagnostics, because of the large spread in oscillator strength values.
Most important in this regard are the Lyman series. A modest S/N coverage
of several of these lines allows for a simple and accurate determination
of $N_{\hi}$. A similar situation occur in the X-ray band where many of
the observed ions are H or He like and therefore show the corresponding
line series.

\section{SUMMARY}

X-ray and UV spectra of NGC~5548 show \ovi\ absorption troughs
associated with the AGN outflow from this galaxy.  A lower limit on
the column density of the \ovi\ X-ray trough is an order of magnitude
larger than the column density found by Brotherton et~al.\ (2002)
using Gaussian modeling of the \ovi\ UV troughs.

It is unlikely that the discrepancy in the inferred $N_{\ovi}$ reflect
actual changes in the \ovi\ absorption between the epochs of the X-ray
and UV observations.  Instead, we interpret the discrepancy as
additional evidence that column densities inferred for UV troughs of
Seyfert outflows are often severely underestimated.  We identify the
physical limitations of Gaussian modeling as the probable explanation
of the $N_{\ovi}$ discrepancy. Specifically, Gaussian modeling cannot
account for a velocity dependent covering fraction, and it is a poor
representation for absorption associated with a dynamical outflow.
Analysis techniques that use a single
covering fraction value for each absorption component suffer from
similar limitations.

\section*{ACKNOWLEDGMENTS}

This work is based on observations obtained with {\em XMM-Newton}, an
ESA science mission with instruments and contributions directly funded
by ESA Member States and the USA (NASA).  Support for this work was
provided by NASA through grant number {\em HST}-AR-09079 from the
Space Telescope Science Institute, which is operated by the
Association of Universities for Research in Astronomy, Inc., under
NASA contract NAS5-26555 and by NASA LTSA grant 2001-029. NA expresses
gratitude for the hospitality of the Astronomy department at the
University of California Berkeley.  The National Laboratory for Space
Research at Utrecht is supported financially by NWO, the Netherlands
Organization for Scientific Research. We also thank Yuri Levin for a
thorough reading of the manuscript.


\end{document}